\documentstyle{article}
\def\be{\begin{equation}}
\def\ee{\end{equation}}
\def\bea{\begin{eqnarray}}
\def\eea{\end{eqnarray}}

\def\br{\biggr}

\def\bl{\biggl}
\def\l{\label}
\def\r{\ref}
\begin{document}

\hfill{}

\vskip 3
 \baselineskip
 \noindent

 {\Large\bf A Dirac Particle in a Complex   Potential}

\vskip \baselineskip

Khaled Saaidi{\footnote {{E-mail-1: KSaaidi@ipm.ir}\\
          {E-mail-2 :   ksaaidi@uok.ac.ir }}}

\vskip\baselineskip

{\small Department of Science,  University of Kurdistan, Pasdaran
Ave., Sanandaj, Iran } \\

 {\small Institute for Studies in Theoretical Physics and Mathematics,
 P.O.Box, 19395-5531, Tehran, Iran}

\vskip 2
\baselineskip


{\bf Keywords}: Complex potential, non-Hermitian, Dirac equation,
energy spectrum.

\vskip 2\baselineskip

\begin{abstract}
It has been observed that a quantum mechanical theory need not to
be Hermitian to have a real spectrum. In this paper we obtain the
eigenvalues of a Dirac charged  particle in a complex static and
spherically symmetric potential. Furthermore,  we study the
Complex Morse  and complex Coulomb potentials.
\end{abstract}

\newpage
\section{Introduction}
 The first interest study in the non-Hermitian quantum theory date back
to an old paper by Caliceti et al \cite{cal}. In \cite{cal} the
imaginary cubic oscillator problem in the context of perturbation
theory has been studied. The energy spectrum of that model is real
and discrete. It
 shows that one may construct  many new Hamiltonians that have
real spectrum, although, their Hamiltonians are not Hermitian. The
key idea of the new formalism (non-Hermitian quantum theory) lies
in the empirical observation that the existence of the real
spectrum need not to necessarily be attributed to the Hermiticity
of the Hamiltonian. In non-Hermitian Hamiltonian, the current
Hermiticity assumption $H^{\dagger} = H$ is replaced by the
PT-symmetry condition as $H^{\dagger} = \hat{P} \hat{T} H \hat{P}
\hat{T} $ [2, 17], where  $\hat{P}$ denotes the parity operator;
$(\hat{P}\psi)(x) = \psi(-x)$  and $\hat{T}$ the time reversal
operator; $(\hat{T}\psi)(x) = \psi^*(x) $  or generally by
Pseudo-Hermitian condition as $H^{\dagger} =\tau^{-1}H\tau$, where
$\tau$ is an invertible Hermitian operator\cite{mostafa1}. Such
non-Hermitian formalism, for the context of Schr\"{o}dinger
Hamiltonian, has been studied  for many different subjects with
several  techniques [1-20]. Also, some explicit studies of the
Hermitian and non-Hermitian Hamiltonians have performed in the
context of Dirac Hamiltonian. For example,
 the
solution of ordinary (Hermitian) Dirac equation for Coulomb
potential including its relativistic bound state spectrum and wave
function was investigated in [22,23]. Also by adding off-diagonal
real linear radial term to the ordinary Dirac operator, the
relativistic Dirac equation with oscillator potential has
 been introduced [24, 25] moreover  the  energy spectrum of
 corresponding  eigenfunctions have been obtaind. The ordinary (Hermitian) Dirac equation for a
charged particle in static electromagnetic field, is studied for
Morse potential \cite{Alhaidari}.

In this paper, we consider the non-Hermitian Dirac Hamiltonian for
complex Morse and complex Coulomb potential. We consider a charged
particle  in static and spherically symmetric four component
complex potential. By applying a unitary transformation to Dirac
equation, we obtain the second order Schr\"{o}dinger like
equation, therefore comparison with well-known non-relativistic
problems is transparent. Using, correspondence between parameters
of the two problems (the Schr\"{o}dinger equation  and the
Schr\"{o}dinger like equation which is obtained after applying the
unitary transformation on Dirac equation for a
 potential ) we can obtain the bound states spectrum and wave function.

The structure  of this article is as follows. In sec.2, we study
the non-Hermitian version of Dirac equation for a charged particle
 with static and spherically symmetric potential, then  by applying  a
unitary transformation we obtain the proper gauge fixing condition
and Schr\"{o}dinger like   differential equation. In sec.3, we
discus the Dirac equation for complex Morse potential then we
obtain  the real energy spectrum and corresponding eigenfunctions.
In sec.4, we consider the  Dirac equation for a complex Coulomb
potential and we  obtain the real eigenvalue of it.

\section{Preliminaries}
The Hamiltonian of a Dirac particle for a complex electromagnetic
field is ($c = \hbar = 1 $)
 \be\l{1} H = \hat{\alpha}. (\hat{p} -
e \hat{A}) + \hat{\beta}m + eV, \ee where the Dirac matrices
$\hat{\alpha} , \hat{\beta}$ have their usual meaning, and setting
$A_{0}$ equal to V. In (1), $\hat{A} $ and V are the vector and
scalar  complex field respectively, where $\hat{A^*} \neq \hat{A}$
and $V^* \neq V$ . Then the Dirac Hamiltonian (1) is not
Hermitian. It is well known that the local gauge symmetry in
quantum electrodynamic implies an invariance under the
transformation as:
 \be\l{2}
 (V, \hat{A})\rightarrow (V' , \hat{A'}) = (V + \frac{\partial
\Lambda}{\partial t}, \hat{A}+ \overrightarrow{\nabla}\Lambda ).
\ee Here $\Lambda(t,\overrightarrow{r})$ is a complex scalar
field. Suppose that the charge distribution is static with
spherical symmetry,  so the gauge invariance implies that $V' = V$
and $\hat{A'} = \hat{r}A(r)$ , where $\hat{r}$ is the radial unit
vector\cite{Alhaidari}. One can denoted the correspondence wave
function of (1)  as: \be\l{3} \Psi=\left (
\begin{array}{rr}
\Phi \\
\chi \\
\end{array}\right ).
\ee In this case  one can obtain
 \bea\l{4}
  (m + eV - E_r )\Phi & =& i{\br [} \hat{\sigma} .
\overrightarrow{\nabla} - e (\hat{\sigma}.\hat{r})A(r){\br ]}\chi, \nonumber \\
 (eV - m- E_r)\chi & =& i {\bl [}\hat{\sigma}.\overrightarrow{\nabla} + e
(\hat{\sigma}.\hat{r}) A(r){\br ]}\Phi. \eea Here $\hat{\sigma}$'s
are the three Pauli spin matrices, $E_r$ is the relativistic
energy eigenvalue,  then we replaced
$ie\hat{\sigma}.\hat{A}(-ie\hat{\sigma}.\hat{A})$ in first(second)
equation of (\r{4}) instead of $e\hat{\sigma}.\hat{A}$,
respectively.   Note that, because of the spherical symmetry of
the complex field, the angular-momentum operator $\hat{J}$ and the
parity operator, $\hat{P}$, commute with the Hamiltonian and the
two spinors $\Phi$ and $\chi$ have opposite parity also. So the
correspondence wave functions are denoted by \bea\l{5}
\Phi &=&  ig(r) \Omega_{\kappa,\mu}(\vartheta , \varphi), \nonumber \\
\chi&=&f(r)\sigma_{r}\Omega_{-\kappa,\mu}(\vartheta , \varphi).
\eea It is seen that \bea\l{6}
 (\hat{\sigma}. \overrightarrow{\nabla} )
 ig(r) \Omega_{\kappa,\mu}(\vartheta , \varphi)&=&
 i\sigma_{r} \Omega_{\kappa,\mu}
 ( \partial_{r}+\frac{1}{r} + \frac{\kappa}{r})g (r),\\
 (\hat{\sigma}.\overrightarrow{\nabla})
 (f(r)\sigma_{r}\Omega_{-\kappa,\mu}(\vartheta , \varphi))&=&
 \sigma_{r}\Omega_{-\kappa,\mu}(\partial_{r} + \frac{1}{r}
 - \frac{\kappa}{r})f(r),
 \eea
where $\hat{\kappa}$ is the spin orbit coupling operator which
defined as: \be\l{8}
  \hat{\kappa}=\hat{\sigma}.\hat{L}+\hbar I.
  \ee
 and we have used from
 \be\l{9}
 \hat{\kappa}\Omega _{\mp\kappa,\mu}(\vartheta , \varphi) =
  \pm \kappa\hbar \Omega_{\mp \kappa,\mu}(\vartheta , \varphi),
 \ee
 in which
\be\l{10} \kappa =\left \{ \begin{array}{rr}
\;\;\;\ {-(l+1)} = -(j + {1 \over 2}) \;\;\;\;\;\; {\rm for} \;\;\;\; j = l +{1 \over 2} \\
l =(j + {1 \over 2}) \;\;\;\;\;\; {\rm for} \;\;\;\; j = l -{1 \over 2} \\
\end{array}\right.
\ee Therefore by defining $u_{1}=g(r)/r$ , $u_{2}=f(r)/r$, we
obtain the following two component radial Dirac equation
\cite{Grainer} \bea\l{11}
 (m + eV - E_r) u_1(r)&=&(\partial_r-\frac{k}{r}-eA(r))u_{2}(r), \nonumber\\
 (eV - m - E_r)u_{2}(r)&=&-(\partial_{r}+\frac{k}{r}+ eA(r))u_{1}.
 \eea
Note that, $A(r)$ is a gauge field, which has a symmetry such as
(\r{2}), therefore, it  must be fixed. It is seen that fixing this
gauge degree of freedom by  $\overrightarrow{\nabla}.
\overrightarrow{A}\equiv \frac{\partial A}{\partial r}= 0$ is not
a suitable choice. Remark that in this paper, instead of solving
Dirac equation we want to solve the $2^{th}$ order differential
equation, which is obtained by eliminating one component of
equation (\r{11}). Note that, this result is not Schr\"{o}dinger
like. One can obtain the proper gauge fixing by applying the
global unitary transformation on two components $u_{1}$ and
$u_{2}$ as: \bea\l{12}
u_1&=&a\phi^u - b\phi^l, \nonumber \\
u_2 &=& b\phi^u + b\phi^l, \eea where $a,b \in \Re$, $a^2 + b^2 =
1 $ and $\phi^u , \phi^l$ are the upper and lower component of
spinor. This unitary transformation  create two results. Firstly,
it makes a gauge fixing condition such as: \be\l{13}
 eV(r)= S (eA(r)+ \frac{\kappa}{r}),
 \ee
where, S= 2ab. Secondly, the final $2^{th}$ order differential
equation is Schr\"{o}dinger like equation. So by applying (\r{12})
on (\r{11}) and use of (\r{13}), we have: \bea\l{14}
 (2eV + Cm - E_r) \phi^{u} + {\bl (}\frac{CeV}{S} - Sm
-\frac{d}{dr}{\br )}\phi^{l}& =& 0,\\
 {\bl (}\frac{CeV}{S} - Sm + \frac{d}{dr}{\br )} \phi^{u} +
 (-Cm - E_r) \phi^{l}&=&0,
 \eea
 where $C= a^{2} - b^{2}$. However, by eliminating  $\phi^{l}$ in
(\r{14}) and (15), one can obtain the Schr\"{o}dinger like
differential equation for radial upper component, $\phi^{u}$, as:
\be\l{16}
 -\frac{d^{2}\phi^{u}}{dr^{2}} + V_{eff} \phi^{u}
 + (m^{2} - E_r^{2}) \phi^{u} = 0, \ee
where
 \be\l{17}
 V_{eff} = (\frac{eC}{S}V)^{2} + (2eE_r V - \frac{eC}{S}\frac{dV}{dr}).\ee

\section{The complex Morse  potential }

The complex Morse potential in Schr\"{o}dinger equation, which
holds discrete spectrum is  given by \cite{Bagchi} \be\l{18}
 V^{CM}(x)= (B_{R} +iB_{I})^{2}e^{-2x} - (B_{R} + iB_{I})(1 + 2D)e^{-x},
 \ee
  the corresponding Schr\"{o}dinger equation is as:
 \be\l{19}
 -\frac{d^{2}\psi}{dx^{2}} + V^{CM}(x)\psi - E\psi =0.
\ee This Schr\"{o}dinger equation is exactly solvable. It is well
known that the energy eigenvalues of the Schr\"{o}dinger equation
for a complex Morse potential (\r{18})  is a function of $(D +
\frac{1}{2})$, which explains that the spectrum of that is real
\cite{Ahmed}, which is obtained as: \be\l{20}
  E_{n} = - (n - D)^{2},
  \ee
where $n$ is an integer number between  0 and  $D$. So we assume
that, the complex potential $V(r)$ in (\r{17}) is:
 \be\l{21}
   V(r) = -(\zeta + i\eta )e^{-r}.
\ee
 Here    $\zeta , \eta  \in\Re$ . By using  equation (17),
one finds the effective complex Morse potential, $V^{CM}_{eff}$,
as: \be\l{22}
    V^{CM}_{eff} = (\frac{eC}{S})^{2} ( \zeta  + i\eta  )^{2} e^{-2r} -
    \frac{Ce}{S}( \zeta  + i\eta  )( 1 -2eE_r) e^{-r}.
\ee By  comparing equations (16) and (22) with (19) and (18),
respectively, we can obtain a  correspondence  between those
parameters as: \bea\l{23}
 \frac{Ce}{S} \zeta &=& B_{R},\nonumber \\
  \frac{Ce}{S}\eta & =& B_{I},\nonumber \\
 -2eE_r &=& 2D, \\
 m^{2}-E_r^{2} &=&-E. \nonumber
  \eea
 Therefore, from (\r{20}) and (\r{23}), it is clearly shown that the real spectrum
 of Dirac particle with complex Morse potential is:
\be\l{24}
 E_{r,n} = \frac{1}{1+e^{2}}{\bl \{}-e n
 + \sqrt{(1+e^{2})m^2-n^{2}}{\br \}}
\ee
 where $ n= 0,1,2,...., n_{max} \leq m\sqrt{1+e^2}$.
 Lastly, we can obtain the eigenfunction of (\r{16}), $\phi^{u}(r)$,  with complex
 effective potential (\r{22}) as:
 \be\l{25}
   \phi_{n}^u(z) = {\cal N}_nz^{\nu_{n}} e ^{-\frac{z}{2}} L_{n}^{2\nu_n}(z), \ee
 where, ${\cal N}_n$ is the normalization constant and
 \bea\l{26}
   z &=& 2 (\frac{Ce}{S})(\zeta + i \eta)e^{-r},\\
   \nu_{n} &=& - eE_{r,n} - n .
   \eea
One can show  that for the case $ n < -eE_{r,n}$, the
$\phi_n^{u}(+\infty) = 0$, and the orthogonality condition is
satisfied  \cite{Ahmed}. Furthermore, from (15), one can obtain
the lower spinor component  as: \bea\l{28}
 \phi_n^{l} (z) &=& \frac{1}{mC + E_{r,n}} (-Sm -
\frac{1}{2} z - z \frac{d}{dz}) \phi_n^{u}(z), \\
& =& -\frac{{\cal N}_n}{mC +E_{r,n}} z^{\nu_{n}}
e^{-\frac{z}{2}}{\bl [}(Sm + \nu_n + n)L_n^{2\nu_n}(z) +(2\nu_n
+n) L_{(n-1)}^{2\nu_n}(z){\br ]},
 \eea
where  (15), (\r{25}), (\r{26}), (27) and recursion properties of
the Laguerre polynomials have been used.

\subsection{The complex Coulomb potential }
We assume that the complex Coulomb potential is: \be\l{30} eV =
\frac{iZ\alpha}{r}, \ee where $\alpha$ is fine structure constant
and $Z$ is atomic number. This equation represents the interaction
of a point charge, -e, and an imaginary point charge iZe. It is
obvious that one has the freedom to use  any unitary
transformation. So for two coupled first order differential
equation (11), with complex Coulomb potential (\r{30}), is
obtained the gauge fixing condition and Schr\"{o}dinger like
differential equation, by applying the another global unitary
transformation, instead of (12). In this case we apply the global
unitary transformation such $e^{\frac{i}{2}\theta\sigma_{x}}$,
which is rewritten as: \be\l{31} U=\left ( \begin{array}{rr}
a & ib \\
ib & a \\
\end{array}\right ),
\ee where $ a , b \in \Re $ and $a^2 +b^2 =1$. By applying (\r{31}
) to $ \phi^{u} $ and $ \phi^{l} $ and institute it in (\r{11}),
we have
 \bea\l{32}
 (m -E_rC)\phi^{u} + {\bl [}\frac{i( S^{2}
- C^{2})}{S} eV - iSE_r - \partial_{r}{\br ]}\phi^l& =&
0, \nonumber \\
{\bl [}\frac{i(S^{2}- C^{2})}{S} eV - iSE_r + \partial_{r}{\br ]}
\phi^{u} - (m + E_r C) \phi^{l} &=& 0, \eea
 where, $S = 2ab$, $C= a^2-b^2$ and we have  used from a gauge fixing
condition as: \be\l{33} eV = \frac{iS}{C}(eA + \frac{k}{r}). \ee
However, we eliminate the $\phi^{l}$ component in (\r{32}), and
obtain the radial differential equation for $\phi^{u}$ as:
\be\l{34}
 - \frac{d^{2}}{dr^{2}} \phi^{u}(r)+
V_{eff}^{CC}(r)\phi^{u} - (E_{r}^{2} - m^{2}) \phi^{u} (r) = 0,
\ee where \be\l{35} V_{eff}^{CC}(r) =
\frac{\gamma(\gamma+1)}{r^{2}} + \frac{2i\alpha Z (S^{2} - C^{2}
)E_{r}}{r} ,\ee in which \bea\l{36}
\gamma &=& \frac{(S^{2} - C^{2})\alpha Z}{S}, \nonumber \\
         &=& \sqrt{\kappa^{2} + \alpha^{2}Z^{2}},\\
         &=&\sqrt{(j + \frac{1}{2})^{2} + \alpha^{2}Z^{2}}
         ,\nonumber
         \eea
is the relativistic angular momentum. Since the wave function has
to be normalized, we have chosen the positive sign for $\gamma $.
Furthermore, it is obviously seen that these solutions, in fact,
exist for all values of Z \footnote{For the ordinary Coulomb
potential $\gamma = \sqrt{k^{2}-Z^{2}\alpha^{2}}$ and, therefore,
one can conclude from $\gamma$ that for states with $\kappa^{2} =
1$ only solution up to $Z \sim 137$ can be constructed. For
$(Z\alpha^{2})> k^{2}$,  in general, the real part of the wave
function shows an oscillatory behavior  and for the states that
$n=\kappa$, the energy spectrum is imaginary \cite{Grainer}.}. The
non Hermitian and PT-symmetric radial Schr\"{o}dinger-Coulomb
differential equation is \cite{Znojil} \be\l{37} {\bl
[}-\frac{d^2}{dr^2} + \frac{l(l+1)}{r^2} + \frac{iA}{r} - E {\br
]} \psi(r) = 0, \ee and its non-relativistic energy spectrum is: ,
\bea\l{38}
 E_{n} &=&  \frac{A^{2}}{4n^{2}},\nonumber\\
  n &=& n_{r} + \gamma + 1,\\
     n_{r} &=& 0,1,2, \ldots \nonumber
     \eea
So comparing equations (\r{34}) and (\r{35}) with (\r{37}), gives
the following correspondences between the parameters of two
problems for the regular solution of $\phi^{u}$ as : \bea\l{39}
 \gamma &=& l,\nonumber\\
 2\alpha Z(S^{2} - C^{2})E_{r} &=& A,\\
(E_r^{2} -m^2) &=& E.\nonumber
 \eea
Using (\r{38}) and (\r{39}) we obtain the relativistic energy
spectrum for complex Coulomb potential as: \be\l{40} E_{r,n} = m
{\bl [ }1- \frac{\alpha^{2}Z^{2}(S^{2}- C^{2})^{2}}{(n_{r} +
\gamma + 1)^{2}}{\br ]}^{-\frac{1}{2}},\ee where $n_{r} = n - j +
\frac{1}{2}$ is the radial quantum number, $n$ the principle
quantum number, $\gamma = j + \frac{1}{2}$ is the angular momentum
quantum number, and consequently, \bea\l{41} E_{r,n} &=& m {\bl
[}1 - \frac{Z^{2}\alpha^{2}(S^{2} - C^{2})^{2}}{[n - j -
\frac{1}{2} + \sqrt{(j + \frac{1}{2})^2 +
Z^2\alpha^2 }]^2}{\br ]}^{-\frac{1}{2}}\\
 n &=& 1,2,3,... \nonumber\\
 j &=& \frac{1}{2}, \frac{3}{2}, \frac{5}{2},....
 \eea
and the upper component of the radial spinor eigenfunction is:
\be\l{43}
   \phi_{n}^u(r) = {\cal N}_nr^{1+\gamma} e ^{-\frac{r}{2}} L_{n}^{2\gamma +1}(r). \ee
Here $r = x -i\theta$, where $x$ is real($x\in [0, \infty)$),
$\theta>0$ and ${\cal N}_n$ is normalization constant
\cite{Znojil}. It shows that the integration path has been shifted
down from the positive. From (\r{35}), it is understood that the
potential is zero in two cases. The first one is  $Z = 0$ and $S
\neq C$, and the second one is  $Z \neq 0, S = C$. It is easily
seen that for vanishing potential the energy
 eigenvalue is $m$.
Note, for $a = \cos(\frac{\pi}{8})$ and $b = \sin(\frac{\pi}{8}),
S = 2ab = \sin(\frac{\pi}{4})=a^{2} - b^{2} =
\cos(\frac{\pi}{4})$. Hence for the case which the unitary
transformation is $e^{\frac{\pi}{8}i\sigma_{x}}$ the energy
eigenvalue is $m$. Therefore, we assume that the unitary
transformation is not identity and also $S \neq C$, then the
vanishing of potential is due to $Z = 0$. Then, from (\r{40}), it
is clearly found that a continuous increase of the coupling
strength $Z\alpha(S^{2} - C^{2})$ from zero, electron states can
be pushed up in to the positive energy continuum. For $Z\alpha
(S^{2} - C^{2})\ll 1$ the energy formula can be expanded as:
\be\l{44} \frac{E_{r,n} - m}{m} \approx \frac{Z^{2}\alpha^{2}(S^2
-C^2)^{2}}{2}{\bl [}\frac{1}{n^{2}} +
\frac{Z^{2}\alpha^{2}}{2n^{3}}(\frac{1}{m} +
\frac{3(S^2-C^2)^{2}}{4n}){\br ]}. \ee It is seen that for
$Z\alpha(S^2-C^2) = 0$ the modified binding energy, $E_{r,n}^{mb}
= E_{r,n} - m$, is \be E_{r,n} ^{mb} = E_{r,n} - m =0, \ee where
$(mb) :\equiv $ (modified binding). Hence, we see that   with
increasing $Z$ the modified binding energy  increases also. Note
that (\r{40}) is nearly real for all values of $Z$ and $n$.
Namely, for the states with $n = j + \frac{1}{2}$, energy value
can be calculated as:
 \be\l{45}
E_{r,n} = m {\bl [}\frac{n^{2} + Z^{2}\alpha^{2}}{n^{2} +
(2CSZ\alpha)^{2}}{\br ]}^{\frac{1}{2}}.\ee Therefore,  this
results show that for all unitary transformations which apply to
Dirac equation for gauge fixing condition and Schr\"{o}dinger like
requirement, the energy eigenvalues of the states with $n = (j +
{1 \over 2})$ is  real for all values of $Z$ (see the second
footnote on the page 7) .

\end{document}